# Steps of the meta-structures project to model general processes of emergence (1)

*Theoretical frameworks, the mesoscopic general vector, future lines of research and possible applications*

Gianfranco Minati
Italian Systems Society, Milan, Italy
Polytechnic University of Milan
gianfranco.minati@AIRS.it

In contrast with classical approaches, we present the project based on considering Collective Behaviours as *coherent sequences of states adopted by different single systems* consisting of the same elements interacting over time in different ways, i.e., through *sequences of variable structures* or phase transitions. This coherence is considered here as being represented by the values taken by suitable mesoscopic variables and their properties represented by Meta-Structures. We introduce a formal tool, i.e., *the mesoscopic general vector* to represent the adoption, over time, of mesoscopic properties by Interacting Collective elements. We explore novel conceptual aspects including *Dynamic* Mesoscopic Levels of Description; necessary and sufficient meta-structural conditions for the establishment and conservation over time of Collective Behaviours; relationships between Meta-Structures and Boundary Conditions to *transform*, even partially, the former into the latter to *prescribe* and induce the emergence of Collective Behaviours as well as coherent Multiple Systems or coherent Collective Beings. We present future lines of research and possible applications.

## INTRODUCTION

This paper is the first of a series about the on-going activity of research related to the Meta-Structures project.

In Section 1 we present approaches and problems when modelling systems using macroscopic and microscopic state variables interacting in *fixed* ways. We then consider Collective Interaction and Collective Behaviour with particular reference to processes of phase transitions, self-organisation and emergence. We then list the approaches used to model and simulate such processes. We mention how structural changes are considered in the literature to model order-disorder transitions.

In Section 2, the main core of the paper, we list some of the fundamental aspects adopted in the project such as the usage of *variable coherent structures*, i.e., variable rules representing the collective behaviour established by elements. This approach is then related to other established approaches in the scientific literature, such as the case of Multiple Systems and Collective Beings and the concept of *order parameter* introduced by Synergetics where the behaviour of a system close to a critical point is considered analogous to that of a system undergoing a *phase transition*. In our approach we consider Collective Behaviour conceptually as *coherent sequences of states adopted by different individual systems* which may exist for any period of time and consisting of the same elements interacting in different ways, i.e. through *sequences of variable structures* or phase transitions. This coherence is considered here as being represented by the values adopted by suitable mesoscopic variables and their Meta-Structural properties. We introduce *the mesoscopic general vector* as a tool for representing the mesoscopic properties adopted over time by collectively interacting elements of a Collective Behaviour. We then provide an overall description of the project as well as related research issues and approaches.

Section 3 deals with possible research issues and approaches.

Section 4 introduces the concept of *Dynamic* Mesoscopic Levels of Description.

Section 5 considers a possible framework for establishing necessary and sufficient meta-structural conditions related to collective behaviours.

Section 6 mentions some relationships between Meta-Structures and Boundary Conditions, as degrees of freedom, on structuring space for interacting agents considered capable of inducing the acquisition of emergent properties in systems. We mention the possibility of *transforming*, even partially, Meta-Structures into Boundary Conditions as well as applying and inducing the emergence of coherent Multiple Systems or coherent Collective Beings. The possibility of adopting the Dynamic Usage of Models (DYSAM), introduced in the literature to model processes of the acquisition of properties in complex systems, may also be considered for modelling multiple Meta-Structures over time.

Section 7 presents future lines of research and possible applications.

Appendix 1 lists some initial measurable variables suggested for Meta-Structural Research.

Appendix 2 lists examples of microscopic, macroscopic and mesoscopic approaches.

Appendix 3 provides definitions of Collective Interaction; Collective Behaviours as established by fixed rules of interacting followed by the component elements; Multiple Systems and Collective Beings; and Collective Behaviours as established by dynamical structures.

## 1. INTRODUCTORY APPROACHES USED TO MODEL AND SIMULATE COLLECTIVE BEHAVIOURS

System behaviour is modelled classically by using interactions between macroscopic variables, as with the Lotka-Volterra equations used to consider the densities of prey and predators in an ecological system to model its evolution. The assumption is that analytical representations of interactions are *fixed* and represented by the invariable equations used with suitable parameters. The approach works for systems which can be described by a limited number of macroscopic state variables and may be formalised using systems of ordinary differential equations. In this view elements are assumed to be *indistinguishable*.

The same analytical approach is also effective for modelling systems established by only a *few* interacting elements such as the solar system. However, the analytical approach is no longer effective when considering microscopic state variables for modelling interactions between a large number of microscopic elements interacting after different initial conditions, applying rules at different times, and in different environmental situations, often simulated by random parameters. However, following the assumption that the elements are indistinguishable, such as particles, and fixing the microscopic interactions over time, some approaches, for studying gases, for example, based on statistical mechanics and statistical physics can be used.

Various methods and approaches have been introduced to study, through simulation, systems established by a huge number of indistinguishable and distinguishable elements interacting through fixed or variable rules, for instance through learning. Such simulations can be based, for instance, on Cellular Automata or Intelligent Agent-Based Models.

### 1.1 Interactions between macroscopic state variables

It is possible to describe the behaviour of a system $S$ by considering suitable macroscopic state variables, e.g., density, pressure or temperature. Instantaneous values adopted by macroscopic state variables $Q_1, Q_2, \dots, Q_n$,, are assumed to specify the state of the system. Behaviour of the system is considered to be represented by the time evolution of the macroscopic state variables, ruled, for instance, by a system of *ordinary differential equation*s as originally introduced by Ludwig Von Bertalanffy (1968, p. 56):

$$\begin{cases} \dfrac{dQ_1}{dt} = f_1(Q_1, Q_2, \dots, Q_n) \\ \dfrac{dQ_2}{dt} = f_2(Q_1, Q_2, \dots, Q_n) \\ \dots\dots\dots\dots\dots\dots \\ \dfrac{dQ_n}{dt} = f_n(Q_1, Q_2, \dots, Q_n) \end{cases} \qquad (1)$$

The system (1) makes evident and specifies how the change in the value of a given macroscopic state variable, affects all the other macroscopic state variables. This represents the concept of *interaction* [1] itself. As specified later, rules explicitly representing interactions between state variables are considered as a *structure* (see Section 2.1.1) of the system. In this view $f_n$ are assumed to be invariable during the process, as is the system (1) representing the structure of the system *S*. Only the values of the variables change over time. At this level of description indistinguishable microscopic elements are allowed to behave in any way, but respecting the values, intended as macroscopic degrees of freedom, adopted by macroscopic state variables as for particles when considering density and pressure.

### 1.2 Interactions between microscopic state variables: Collective Interaction and Collective Behaviour

The above approach may also be used to model interactions between microscopic elements by considering microscopic state variables when dealing with only a few, fixed rules as for a pendulum, oscillators or the solar system.

However, when modelling a number $N \geq 3$ of collectively interacting indistinguishable elements, i.e., adopting the so-called homogeneous approach, analytical representations and solutions may still exist, although found only with great difficulty for particular cases. Jules Henri Poincaré (1854-1912) introduced the so-called *three-body* problem discovering a chaotic deterministic system which laid the foundations of modern chaos theory. Researchers had to analytically model 6*n* variables, since each particle is represented by three positional and velocity components. In physics there is the well-known *N-body* problem studied by classical mechanics consisting of finding, given the initial positions, masses, and velocities of *N-bodies*, their subsequent motion. The problem is complicated when particles interact according to different rules, as for gases and electrons when the influence of each particle upon every other can be represented by an inverse square law, such as Newton's law of gravitation, or Coulomb's law of electrostatic interaction. The *N-body* problem has been thoroughly studied in the literature using various approaches (Aarseth *et al.*, 2008). These include the use of the methods of statistical mechanics and statistical physics, chaos and stability analysis, and *N-body* algorithms for simulations. Elements are considered as particles and modelled, *all* following the *same* behavioural rules, fixed over time. The *large number* of cases due to: 1) elements interacting starting from different initial conditions; 2) frequencies, combinations, intensity, angularities of random sequences or eventually simultaneous, i.e., parallel, input; 3) different rules of interaction, such as gravitational and electrostatic; 4) different environmental situations, often simulated by random parameters; do not allow classic analytical representations to be separately computed for each configuration over time. This is what it is intended by *Collective Interaction*. The focus is placed upon modelling the microscopic behaviour of single particles.

An extension to this should also be considered where rules are *variable* and due, for instance, to *evolutionary* processes or processes of learning performed by elements considered as agents, possibly possessing cognitive systems, e.g., considered as computational systems able to learn. This is the case for heterogeneous approaches when elements process input in different ways due to different contextual, environmental situations or properties of memory, language and emotions eventually establishing an, even simulated, cognitive system. This is the case, e.g., for traffic, crowds, swarms and flocks. Simulations are performed, for instance, using Agent-Based Models, computational models of Multi-Agent Systems simulating the actions and interactions of *autonomous elements* (Lane *et al.*, 2009). Simulations are based upon the combination of a huge variety of approaches and tools such as genetic algorithms for evolutionary programming, Neural Networks, Cellular Automata, Game Theory, computational algorithms as repeated random sampling, such as Monte Carlo Methods used to introduce randomness.

[1] This classical definition of interaction should be further elaborated within the framework of new concepts such as those of QFT and the related revision of the concept of particle, briefly discussed below.

If Collective Interaction establishes a collective entity detected by the observer, at a suitable level of description, having properties which the component elements do not possess, we may distinguish (Minati, 2008a) between processes of:

a) phase-transition corresponding to the acquisition of, or a change in, structure, as for first and second-order phase-transitions, e.g., water-ice-vapour; paramagnetic-magnetic and conductivity-superconductivity;
b) self-organisation corresponding to continuous but *stable*, for instance, periodic, quasi-periodic and predictable, variability in the acquisition of new structures, as for Bènard rolls, the Belousov-Zhabotinsky reaction, coherence in light emission typical of the laser, swarms having repetitive behaviour, and dissipative structures, such as whirlpools in the absence of any internal or external fluctuation. Stability of variability, e.g., periodicity, corresponds to stability of the acquired property;
c) emergence corresponding to continuous, irregular, but *coherent*, i.e., as detected by the observer using a suitable cognitive model and related level of description, variability in the acquisition of new structures, as for swarms and flocks adopting variable behaviours in the presence of any suitable environmental condition, or Industrial Districts. *Multiple* and *subsequent* systems are not hierarchical, but sequential and coherent over time, i.e., they display to the observer the *same* emergent, acquired property. This is the case for the emergence of Collective Behaviours.

In the latter case (Baas, 1994; Baas and Emmeche, 1997) emergent macroscopic acquired properties are modelled as generated by coherence between systems of microscopic collective interactions.

The theoretical role of the observer is to adopt a level of description and cognitive model suitable for detecting emergent properties. The observer may not only *possess*, but eventually *acquire* new levels of description and also design artificial observers able to use specific levels of description. We may consider, for instance, levels of description for processes of vision adopted by the observer, such as temporal, syntactical and semantic (Licata, 2008a) as well as the use of memory, whether working, episodic, or semantic. It is not a question of *relativism*, but rather of *constructivism* (Butts and Brown, 1989; Von Glasersfeld, 1995) as is the usage of Multiple Models to deal with processes of emergence of multiple acquired properties in complex systems (Licata, 2007a; Minati and Pessa, 2006).

The *many-body problem* is now a version of the *N-body* problem studied in quantum mechanics, i.e., at another theoretical level of description (Licata, 2007b; Schlosshaue, 2007).

The purpose of current research on emergence is to model the emergence of Collective Behaviours by using a language able to represent *both* the phenomenon under study *and* the process of observing or the observer as an active *cognitive generator* of the phenomenon. This relates to the *coupling* of the phenomena and the cognitive model used by the observer as in a configuration of mutual observers modelled by *logical openness* (Minati, Penna, Pessa, 1998; Licata 2008b) and DYSAM (Minati and Brahms, 2002; Minati and Pessa, 2006, pp. 64-75). Current research is still based on the conceptual *separation* of the observed phenomenon from the observer as well as the existence of localizable particles possessing properties. Our approach is based upon the constructivistic *invention* of a suitable mesoscopic level of description by the observer. Selection and invention of the mesoscopic level of description is assumed to *represent* interests, conceptual frameworks, problems, levels of knowledge, and purposes of the observer as introduced in Section 2.1.7.

This refers to the concept of *intrinsic emergence* proposed by Crutchfield (Crutchfield, 1994) "…referring to a process in which the occurrence of a certain behaviour is not only unpredictable, but its occurrence gives rise to profound changes in system structures such as to require a formulation of a new model of the system itself" (Minati and Pessa, 2006, p. 96).

However, this subject is currently under discussion in the scientific community when distinguishing between *phenomenological/computational emergence* and *radical/observational* *emergence*. This is based on a criticism of the Baas-Emmeche definition of emergence introduced by the physicist P. W. Anderson's famous statement 'more is different' (Anderson, 1979). When dealing with *phenomenological/computational emergence* it is assumed that is always possible to find a formal description of the relation between the lower and the emergent level, e.g., though simulation models. On the contrary, *radical/observational emergence* is considered to occur when it is impossible to find such a relation. This is the case considered by P. W. Anderson (Anderson, 1981; Anderson and Stein, 1985) in focusing upon processes of *spontaneous symmetry breaking* occurring in Quantum Theory, e.g., superconductivity and superfluidity. Such a description of emergent phenomena is possible *only* within the theoretical framework of Quantum Theory (Sewell, 1986). The problem of adopting such a strong position relates, in our view, to the difficulties in representing and modelling phenomena using the QFT formalism, such as processes of phenomenological emergence at different levels of descriptions as for learning, social systems, e.g., revolutions, and collective behaviours.

### 1.3 Modelling and Simulating

Various approaches for simulating the emergence of Collective Behaviours have been introduced in the literature (see Minati and Pessa, 2006). Some are based upon the microscopic approach, i.e., by considering individual agent behaviours, including:

a) Rule-based

- separation rules: individuals must control their motion in order to avoid the crowding with adjacent components;
- alignment rules: individuals must control their motion so as to point towards the average motion direction of adjacent components;
- cohesion rules: individuals must control their motion so as to point towards the average position of adjacent components.

Examples of rules are:

1. max distance < M;
2. min distance > m;
3. distances always change with time;
4. different directions among agents, but within an angle < α.

b) Lattice models

Individual agents are represented as moving entities localized within a discretised spatial lattice. The motion of the single agents and of the system of interacting agents is given by suitable local evolutionary rules, giving the state of a lattice point at a given point in time as a function of the state of the neighbouring points at the previous point in time (e.g., CA).

c) Hydrodynamic-like models

This term refers to models based on a macroscopic continuum-like description of a system of agents, viewed as components of a suitable *fluid*. The emergence of Collective Behaviour is identified by the appearance of large-scale spatially dependent stationary stable or metastable solutions of the macroscopic evolution equations. The advantage of this approach is that it makes explicit the factors controlling the onset and the conservation of these "collective" solutions. These factors include:

- local inhibition: this comes from the fact that the operation of each single cognitive system requires the processing of information arriving from external stimuli; the information to be processed may be assumed to come from adjacent individuals; however, such information processing requires a significant amount of energy; it may therefore be viewed as equivalent to the effect produced on individuals by a force, opposing individual movement; such a force, then, may be considered as dependent only upon nearest neighbours;

- medium range activation: this comes from the very existence of a tendency towards cooperation or behavioural imitation between different agents; it is medium range, owing to the fact that the perceptual abilities of every agent can cover only a limited spatial range, although greater than the shorter range between nearest neighbours;
- long range inhibition: this is due to the existence of forces which counteract the control abilities of each agent (e.g., friction); typically their effect manifests itself only over very large spatial and/or temporal scales.

Within this conceptual framework Collective Behaviours emerge and are maintained as a consequence of a suitable balance between these various factors (Minati and Pessa, 2006, pp. 104-105).

We may also conceptually consider the cases given by

- A network of oscillators in which the output of one is processed by those connected to it. However, connections may be *weighted* by introducing resistors and capacitors. The network constitutes the structure of the system and it is assumed as being *variable* over time changing both connections and weights.
- A Neural Network having a *variable* structure and architecture, e.g., weights and layers between neurons.
- A Cellular Automata having *variable* evolutionary rules.

Collective Behaviours occur, i.e., the emergence of a property such as sequences of patterns, when variations are *coherent*. The cases above are assumed to represent coherent processes of *subsequent phase transitions*, i.e., changes in structure.

Regularity and periodicity in the variations of structures represent processes of self-organisation.

## 2. THE PROJECT

The purpose of this project is to introduce Meta-Structures as a way of formally modelling the concept of coherence as related to the emergence of acquired properties from collectively interacting elements, such as stable entities like swarms and flocks. As we will see the constructivistic theoretical role of the observer is to identify suitable variables to represent properties and the process. We recall, as in physics, for instance, the concept of the coherence of two waves relates, on the contrary, to how well correlated they are as quantified by the cross-correlation function quantifying the possibility of predicting the value of the second wave by knowing the value of the first. We have *self-coherence* when the second wave is not a separate one, but the first wave at a different time or position.

We will consider coherence as coherent variations in structure. This phenomenon is too complicated to be treated analytically and thus we propose the use of Meta-Structures.

### 2.1 Outlines and approaches adopted

#### 2.1.1 Organisation and Structure

In the systemics literature, following the contributions of Maturana and Varela (Varela *et al.*,1974; Maturana and Varela, 1980) and in contrast with mathematics and engineering, researchers adopted the following definitions. While *organizations* deal with networks of relationships having undefined parameters, *structures deal with* networks of relationships having well-defined parameters. A specific structure is one among several possible *applications*, or *specifications* of an organisation suitable for making the system acquire desired properties or modelling a given phenomenon in this way. One example is the distinction between organization and structure describing a Neural Network. Its organisation may be constituted of *n inputs*, *m hidden layers* and *s outputs* while its structure is given by a network with precise values of connection weights and well-defined transfer functions associated with individual neurons. Organisation relates to the general architecture of a system, whereas a structure relates to a specific type of system having this organisation. Structure is intended as given by rules of interaction between state variables. Equations modelling the evolution of Dynamical Systems, e.g., the Brusselator, Lorenz or Lotka-Volterra equations, may be considered as *suitable structures setting relationships between state variables.* Rules used for evolving, rule-based systems, e.g., Cellular Automata and Agent-Based Models, may be considered as specific structures when using specific rules, setting constraints upon elementary interacting elements, i.e., cells and agents.

Structure is consequently considered (Minati, 2008a) through the rules of interaction between single elements represented by a) microscopic state variables dealing, for instance, in flocks and swarms, with relationships between mutual separation distances, speed, direction, and altitude or b) macroscopic state variables dealing, for instance, with relationships between pressure, temperature and density. In these cases structures between microscopic or macroscopic variables are assumed as *fixed*, i.e., the analytical representations or rules do not change. *The dynamics are given by the interaction itself and not by the changing of the ways in which interaction occurs*.

### 2.1.2 Dynamic Structures

In our project we consider the structures between microscopic or macroscopic variables as *variable*, i.e., the analytical representations of rules change over time as well the elements to which those rules are applied. *Dynamic structures* are assumed to change over time in such a way as to cause Collective Behaviour to emerge from Collective Interactions, i.e., dynamic structures change in a *coherent* way. The approach is similar to that mentioned above for modelling processes of phase transition, self-organisation and emergence based on considering structures, their variability and coherence (see Section 1.2).

This process of the changing of structures between microscopic or macroscopic variables may ideally be studied by considering families of functions changing over time. This may be represented in particular cases by periodic and quasi-periodic functions [2], and more generally by functions defined, for instance, by parameters and exponents depending upon time or by families of functions defined as corresponding to single points in time $t_i$. However such an *analytical* representation is very difficult as is the case for the *N-body* problem. A common strategy is based upon modelling and simulating by using sub-symbolic approaches based, for instance, upon Neural Networks, Cellular Automata and Agent Based Models, avoiding any analytical representation.

### 2.1.3 From structural change to meta-structural change

It is well-known that some processes of change, such as *second-order phase transitions*, are characterized by the absence of a difference in the density of the two phases and of latent heat emitted or absorbed in the transition. Such a transition occurs abruptly and simultaneously within the whole system under consideration. At the critical point there is no co-existence of the two phases, the new and the old one, but the transition from one phase to the other. This kind of transition consists of an internal rearrangement of the *system structure*, occurring at the same time at all points within it. In other words, the transition occurs because the conditions necessary for the stable existence of the structure corresponding to the initial phase *cease to be valid* and a new stable structure replaces it. Examples are transitions from a paramagnetic to a ferromagnetic state, the occurrence of superconductivity and of superfluidity, and order-disorder transitions in some kinds of crystals.

---

[2] A function $h:R \rightarrow R^n$ is considered *quasi-periodic* if it can be written as $h(t)= H(\omega_1 t, \ldots, \omega_n t)$, where $H$ is periodic with a period $p = 2\pi$ in each of its arguments and two or more of the $n$ frequencies are incommensurable, i.e., the ratio between two or more frequencies is a non -rational number, see Bruno (1995). Examples of quasi-periodic phenomena are illustrated by Keller (1999) and Hemmingsson and Peng (1994).

We may consider the *dynamics of a single system* when suitable macroscopic state variables adopt sequences of different values over time. In organised systems, such as mechanical systems, electronic devices and assembly lines, systemic properties are established by components interacting while respecting a *specific* structure. Properties of organised systems are modelled by suitable macroscopic state variables. Macroscopic state variables acquire *different values* during the process of change like levels of specific systemic functionalities for electronic devices. In this case, the dynamics are given by sequences of values taken by macroscopic state variables over time.

We may also consider the *dynamics as sequences of various single systems* established over time by the same elements interacting in different ways, i.e., having variable structures, as introduced in Sections 2.1.4 and 2.1.5. Emergent systems, such as flocks, swarms, traffic and Industrial Districts, may indeed be considered as coherent Multiple Systems or coherent Collective Beings (see Section 2.1.5), or established, from a more general point of view, by *coherence* between *sequences of states adopted by various single systems* which may exist for any period of time and consisting of the same elements interacting in different ways, i.e., by *sequences of variable structures*. The coherence of these dynamics, making emergent one or more acquired properties, is proposed here suitably modelled by using Meta-Structures. **An emergent system is considered to be established by *coherent sequences of states adopted by various single systems* consisting of the same elements interacting in different ways, i.e. by *sequences of variable structures* and not as *sequences of states of the same system.*** An emergent system acquires emergent properties such as shape and behaviour.

| Organised systems, (electronic devices) | Emergent systems, (flocks and traffic) |
|---|---|
| The dynamics are due to sequences of values related to different states of the same system over time (same structure) | The dynamics are due to coherences between *sequences of states adopted by different single systems* which may exist for any period of time and consisting of the same elements interacting in different ways over time which we consider as corresponding to *sequences of different structures* (different structures) |

Table 1: Dynamics of systems

### 2.1.4 Modelling Collective Behaviours by using Dynamic Structures

In our project we adopt the conceptual framework of Collective Interaction *and* the working hypothesis that different rules affect the microscopic behaviour of different elements in a *non-homogeneous* way.

More specifically, we consider:

a) a set of *m* elements $e_k$ with $3 \leq k \leq m$, finite and limited;
b) a set of only a *few* behavioural rules $r_s$ for elements $e_k$ , such as those listed in Section 1.3. We assume that several rules may apply to the same element and the same rule may apply to several elements;
c) a finite discrete observational time, i.e., computational cycles $t_{i:1,w}$ where $1 \leq i \leq w$, finite and limited;

d) The set $St_i$ consisting at time *i* of *all* subsets $St_i(r_s)$ of the elements $e_k$ interacting according to rules $r_s$. At time $t_i$, rules $r_s$ may apply to elements $e_k$ in any combination or way. Sets $St_i$ may have elements in common;
e) The set *S* consists of *all* sets $St_{i:1,w}$

In this view a Collective Behaviour in 2D may ideally be represented over time by sequences of virtual sets of different instantaneous superimposed independent layers, i.e., *S* over time, where common elements eventually interact by following single instantaneous different independent, but fixed rules. Moreover, elements may simultaneously belong to different virtual layers, i.e., interact by simultaneously following different fixed rules.
In 3D a Collective Behaviour may correspondingly be represented over time by different instantaneous *crossing* of independent virtual sequences of clusters of elements interacting according to single instantaneous different independent but fixed rules. In an analogous way, elements may simultaneously belong to different virtual clusters, i.e., interact by simultaneously following different fixed rules.
This is the case introduced in the literature and mentioned in Section 2.1.5 when sequences of subsets or clusters of the *m* elements $e_k$, considered at point d) above establish coherent Multiple Systems or coherent Collective Beings by different simultaneous multiple roles as at the above point e), see Section 2.1.5 (Minati and Pessa, 2006).
This case is further generalised when Collective Behaviour is considered as being established by coherent states adopted by sequences of systems established by variable, virtually infinite, behavioural rules $r_s$. Processes of emergence of Collective Behaviours are proposed to be modelled through the use of suitable mesoscopic clusters of elements adopting the *same* values of some microscopic state variables and related general properties over time, see Sections 2.1.6 and 2.1.7.

### 2.1.5 The case of Multiple Systems and Collective Beings

We recall that a Multiple System is a set of simultaneous or successive systems modelled by the observer and established by the same elements interacting in different ways, i.e., having multiple roles simultaneously or at different times (Minati and Pessa, 2006, pp. 89-134). Successive systems consisting of the same elements interacting in different ways over time and establishing a Multiple System may exist for any period of time.

Examples of Multiple Systems are electricity networks where different systems play different roles in continuously new, emerging usages allowing emergent properties such as a black-out or networked interacting computer systems performing cooperative tasks as on the Internet. The study of Multiple Systems also considered *interchangeability* between interacting agents to model emergent behaviour as ergodic, see Section 2.1.7.3. As we will see at the Section 2.1.5.1, Multiple Systems as simultaneous and subsequent systems established by the same interchangeable (as represented by equations 6, 7 and 8) elements interacting in different ways are *coherent* when suitable constraints simultaneous control the validity of various interactions such as those expressed by equations (9). Coherence corresponds to the acquisition of an emergent property by the Multiple System.

Collective Beings are particular cases of Multiple Systems when elements are autonomous agents all provided with the *same* cognitive system and they may *decide*, within their physical and cognitive limits, upon their way of interacting. When considering human Collective Beings, these are established by using *different cognitive models*. Examples are cases where elements may *simultaneously* belong to different systems (e.g., components of families, workplaces, traffic systems, consumers, and mobile telephone networks) and *dynamically*, i.e., at different times, giving rise to different systems, such as temporary communities (e.g., audiences, queues, and passengers on an airplane). Section 2.1.5.1 recalls that, in the original book cited above, the source of *coherence* between simultaneous or successive Collective Beings is considered as the belonging of interacting agents to the same species which thus possess the same cognitive system and the same physical characteristics, i.e., degrees of freedom.

Multiple Systems and Collective Beings are established simultaneously and at different times by the same elements interacting in different ways, i.e., by multiple simultaneous or successive structures.

In the first case, i.e., simultaneity, multiple structures apply at the same time to the same or different elements. In this case Multiple Systems and Collective Beings are established by simultaneous systems.

In the second case, i.e., at different times, multiple structures apply at different times to the same or different elements. In this case Multiple Systems and Collective Beings are established through sequences of systems, see Table 2.

*Sequences of states adopted by corresponding sequences of various single systems* established over time by the same elements interacting in different ways, i.e., having variable structures, establish coherent Multiple Systems or coherent Collective Beings when their coherence is due to multiple and interchangeable roles, such as speed, altitude, distance, topological position in a flock of boids, and constraints such as those expressed by equations (9) or to usage of the same cognitive system to interact.

As seen in Section 2.1.6, the adoption of coherence by *sequences of states adopted by corresponding sequences of various single systems* established over time by the same elements interacting in different ways, i.e., having variable structures, is further generalised by considering the adoption of suitable Meta-Structural properties. Indeed, constraints such as those establishing coherent Multiple Systems or coherent Collective Beings are expected to be suitably transformed into equivalent Meta-Structural properties.

Processes of self-organisation and emergence may be modelled as *coherent sequences* of *states adopted by different single systems* which may exist for any period of time and consisting of the same elements interacting in different ways over time.

In a particular case *different single systems* may be the ones establishing Multiple Systems and Collective Beings. In case of self-organisation coherence for Multiple Systems may be due to periodicity and quasi-periodicity of systems, i.e., structures, see Section 1.2, point (b.

As shown in Section 2.1.7, roles considered by the observer to detect and model such coherence are not arbitrarily determined but, rather, based on cognitive *Gestalt continuity*, extensions or replications of conceptual categories used to model normal non-collective behaviours. We refer, for instance, to the ‘good continuation’ and ‘imitation’ principles assuming that the perception of a configuration includes the imaginable process by which the configuration was assumed to have been created (Arnheim, 1997; Bongard, 1970). When considering a flock of birds, roles may then be, for instance, maintaining altitude, speed, topological positions, direction and distance.

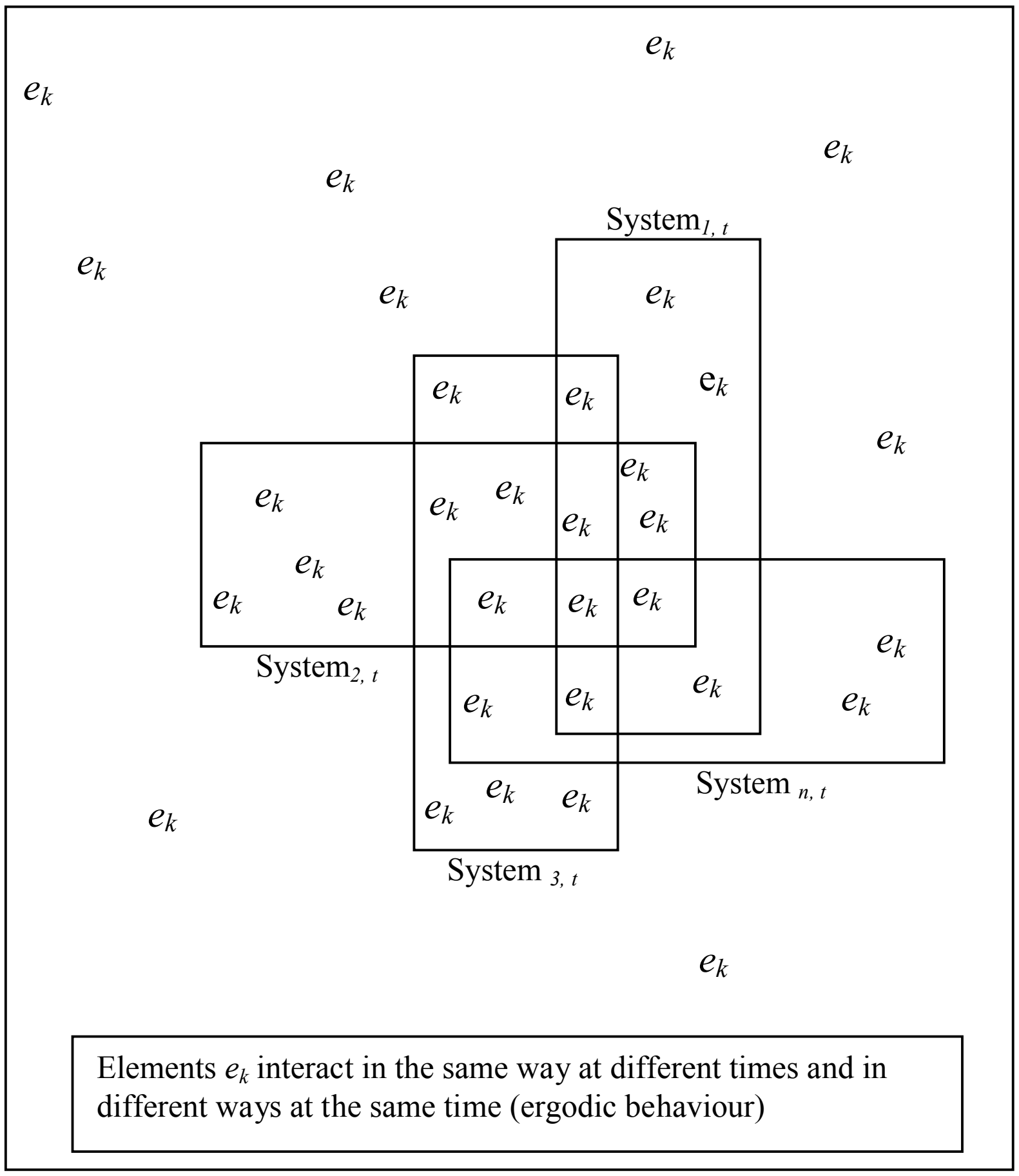

Table 2: elements $e_k$ establishing multiple, possibly different systems over time through multiple simultaneous interactions

For a proposed formalisation of the concepts of Multiple Systems and Collective Beings we refer to the approach presented in Minati and Pessa (2006, pp. 123-128). Starting from the framework of dynamical systems theory, consider a conceptual formalisation based on autonomous systems of differential equations. Consider, for instance, three systems, $S_1$, $S_2$, and $S_3$ , each defined by a suitable set of macroscopic state variables $v_i(t)$ and that some are common to more than one single system. The evolution of macroscopic state variables $v_i(t)$, representing system behaviour, is modelled, for instance, by a system of *fixed* ordinary differential equations. We may consider, for instance, the following case:

$$S_1: (v_1, v_2, v_3, v_5, v_6, v_7)$$

$$S_2: (v_4, v_5, v_6, v_8) \qquad (2)$$

$$S_3: (v_6, v_7, v_9)$$

A Multiple System over time may be described by *n*, in this case three, different systems of autonomous differential equations simultaneously valid and representing the dynamical and simultaneous roles of some elements, such as:

$$
S_1: \begin{cases} dv_1/dt = f_1(v_1, v_2, v_3, v_5, v_6, v_7) \\ dv_2/dt = f_2(v_1, v_2, v_3, v_5, v_6, v_7) \\ dv_3/dt = f_3(v_1, v_2, v_3, v_5, v_6, v_7) \\ dv_5/dt = f_5(v_1, v_2, v_3, v_5, v_6, v_7) \\ dv_6/dt = f_6(v_1, v_2, v_3, v_5, v_6, v_7) \\ dv_7/dt = f_7(v_1, v_2, v_3, v_5, v_6, v_7) \end{cases} \qquad (3)
$$

$$
S^2: \begin{cases} dv_4/dt = f_4(v_4, v_5, v_6, v_8) \\ dv_5/dt = f'_5(v_4, v_5, v_6, v_8) \\ dv_6/dt = f'_6(v_4, v_5, v_6, v_8) \\ dv_8/dt = f_8(v_4, v_5, v_6, v_8) \end{cases} \qquad (4)
$$

$$
S^3: \begin{cases} dv_6/dt = f''_6(v_6, v_7, v_9) \\ dv_7/dt = f'_7(v_6, v_7, v_9) \\ dv_9/dt = f_9(v_6, v_7, v_9) \end{cases} \qquad (5)
$$

It should be noted that the common state variables, in this case $v_5$, $v_6$, and $v_7$, *simultaneously* behave as components of different systems, as in the following evaluations:

$$
\begin{cases} dv_5/dt = f_5(v_1, v_2, v_3, v_5, v_6, v_7) \\ dv_5/dt = f'_5(v_4, v_5, v_6, v_8) \end{cases} \qquad (6)
$$

$$
\begin{cases} dv_6/dt = f_6(v_1, v_2, v_3, v_5, v_6, v_7) \\ dv_6/dt = f'_6(v_4, v_5, v_6, v_8) \\ dv_6/dt = f''_6(v_6, v_7, v_9) \end{cases} \qquad (7)
$$

$$
\begin{cases} dv_7/dt = f_7(v_1, v_2, v_3, v_5, v_6, v_7) \\ dv_7/dt = f'_7(v_6, v_7, v_9) \end{cases} \qquad (8)
$$

Equations considered for modelling the coherences of sequences of Multiple Systems of this example may be described by set of functional constraints representing the simultaneous validity of equations (6), (7), and (8) above.

An example of functional equations representing functional constraints is:

$$
\begin{cases}
f_5\,(v_1, v_2, v_3, v_5, v_6, v_7) = f'_5\,(v_4, v_5, v_6, v_8) \\
f_6\,(v_1, v_2, v_3, v_5, v_6, v_7) - f'_6\,(v_4, v_5, v_6, v_8) = f'_6\,(v_4, v_5, v_6, v_8) - f''_6\,(v_6, v_7, v_9) \\
f_7\,(v_1, v_2, v_3, v_5, v_6, v_7) = f'_7\,(v_6, v_7, v_9)
\end{cases}
\tag{9}
$$

Note how the functional equations (9) introduce constraints by reducing the number of degrees of freedom of the original description based upon the systems of equations (3), (4) and (5).

On the one hand, the representation of simultaneous equations (3), (4), (5) requires a 9-dimensional phase space, while on the other, the assumption of the three constraints expressed by (9) allows for an evolution in only a 6-dimensional phase space.

The reduction of the number of degrees of freedom has significant implications regarding the stability of the motions of the whole system. It is related to the fact that, on increasing the dimensionality of the phase space the number of ways in which an equilibrium state can become unstable also increases. For instance, an equilibrium point, stable when considered in two dimensions, could be unstable when considered in three dimensions, e.g., a spiral motion towards an equilibrium point in two-dimensional space could be simply the two-dimensional shadow of helicoidal motion wandering away from that point along the third dimension.

By adopting the inverse notion, a reduction in the number of degrees of freedom increases stability, e.g., representing a spiral motion on a two-dimensional plane, allows one to escape the loss of stability occurring in the third dimension.

This is the reason why a Collective Behaviour, like a Collective Being, is, in principle, more stable than its local constituent parts, this stability being granted by defining constraints. This explains why some 2D collective behaviours seem to violate well-known theorems of Physics, such as the Mermin-Wagner theorem stating that a stable two-dimensional configuration cannot exist (Mermin and Wagner, 1966). The reason is that 2D collective behaviours are a consequence of suitable constraints between the motions of individual components, lowering the dimensionality of the available phase space. This increases the stability of the whole system and renders inapplicable the thermodynamic arguments upon which the Mermin-Wagner theorem is based.

### 2.1.5.1 Coherences in Multiple Systems and Collective Beings

In the case of self-organisation, the coherence of sequences of states adopted by Multiple Systems is due, for instance, to periodicity or quasi-periodicity of such systems, i.e., structures, as for Bènard rollers, Belousov-Zhabotinsky reactions, the laser, repetitive collective behaviours, e.g., swarms, and dissipative structures such as whirlpools, when the non-perturbed. Persistence of coherence corresponds to a stability of the acquired property. In this case fixed rules as in equations (9) are substituted by fixed rules of periodicity or quasi-periodicity. Systems composing a coherent Multiple System may have different temporal duration as well the Multiple System itself, since interest focuses upon the coherence of sequences of states adopted. A coherent Multiple System may temporally coincide with one of its composing systems.

In the case of Collective Beings, the coherence of sequences of states adopted by Multiple Systems is not due to suitable fixed rules as in equations (9), nor to their periodicity or quasi-periodicity, but rather *prescribed* by the *same* cognitive system and the *same* physical characteristics used by autonomous agents to interact, being assumed belong to the *same biological species* as for swarms and flocks.

#### 2.1.5.2 Collective variables

In our generalised approach, introduced in Section 2.1.6, for modelling general Collective Behaviours we move from the description of Multiple Systems based on the systems of equations (3), (4), (5) to a new description, based on suitable *collective variables*, mesoscopic in our case, with which collective properties can be described. Coherence establishing Collective Behaviour will be considered here as modelled by Meta-Structures as properties of suitable mesoscopic variables.
The introduction of collective variables is a widely used tool in theoretical physics, allowing one to move from representations of a system based, for example, upon a set of isolated atoms, considered as mutually interacting in a very complicated way, to a new collective representation (physically equivalent to the previous one) based on isolated atoms interacting in a simple way, but with suitable collective excitations, as in the case of the so-called *quasi-particles*[3].
We may also notice that the macroscopic level of description does not allow one to suitably model *transient processes*, such as the establishment, desegregation, merger and splitting of Collective Behaviours. We will see this can be modelled by changing the Meta-Structural properties.

### 2.1.6 The generalised approach

In order to introduce a suitable approach for modelling general processes of the emergence of Collective Behaviours, i.e., the acquisition of coherence between component behaviour over time, we propose to *model* the coherence of sequences of systems establishing Collective Behaviour by using the properties of various, and possibly simultaneous mesoscopic variables.
Within this conceptual framework we consider coherence between sequences of such individual systems as represented by suitable mesoscopic clusters of elements adopting the *same* values of some suitable microscopic state variables and related Meta-Structural properties over time, rather than fixed constraints such as those represented, for instance, by equations (9). We may conceptually consider Meta-Structural properties as variable constraints suitable for inducing the components to which they apply, i.e., microscopic or macroscopic variables, to acquire coherent behaviours. **In this conceptual framework fixed constraints, such as those represented by equations (9) establishing coherent Multiple Systems, should be *transformable* into suitable Meta-Structures.**
We consider *instantaneous, subsequent values* adopted over time by suitable mesoscopic variables, such as the number of elements at the *same* distance, see Section 2.1.7.1, and

- properties of parameter values defining the mesoscopic variable, such as the distance considered, i.e., Meta-elements, see Section 2.1.7.2; and
- mathematical properties, i.e., Meta-Structures, possessed by temporally ordered sets of values adopted over time by Mesoscopic state variables and Meta-elements, see Section 2.1.7.3.

Although structures relate to interacting elements, meta-structures are considered to model the coherence of non-regular variable structures by considering sequences of values adopted by mesoscopic state variables and their related parameter Meta-elements.
Within a more general framework we consider *any* sequence of simultaneous, multiple systems to establish collective, i.e., coherent, behaviour when respecting suitable Meta-Structural properties.

---

[3] Quasi-particles share with traditional particles many features, except localization.

**Thus, although any coherent Multiple System or coherent Collective Being is assumed to be modelled by using Meta-Structures, not *any* Collective Behaviour modelled using Meta-Structures is established by coherent Multiple Systems or coherent Collective Beings.**

The process of the establishment of coherent Multiple Systems and coherent Collective Beings is a particular case, as constituted by sequences of systems consisting of ergodic percentages of conceptually interchangeable agents taking on the *same roles at different times and different roles at the same time*, or governed by constraints as mentioned above. This case is considered as being represented in a more general way by Meta-Structural properties.

However, in the general case under study, coherence, as detected by the cognitive model of the observer, may be of *any* kind, i.e., not related, for instance, to multiple, different roles of interchangeable agents. There are no *formal* rules, e.g., stating continuity or coherence, between variable structures. Structures may involve, in different ways over time, different elements.

In the general case under study coherence is given by keeping an *emergent* property adopted by sequences of systems such as shape and collective learning abilities in social systems, flocks, swarms, markets, industrial districts, traffic, and functionalities in networks of computers (e.g., on the Internet). Indeed, in the case of Multiple Systems and Collective Beings, systems are coherent due to multiple simultaneous roles of elements and their interchangeability, while in the case of general Collective Behaviours their coherence over time is considered as being represented by Meta-Structures.

| Collective Behaviours as coherent Multiple Systems/coherent Collective Beings | General Collective Behaviours modelled by Meta-Structures |
|---|---|
| Set at time $t_i$ of different simultaneous or successive coherent systems established by the same elements interacting in different ways, i.e., having multiple simultaneous or dynamical (different times) roles such as *multiple and simultaneous phase transitions*, see Table 2. Collective properties are due to continuity and coherence of the multiple dynamic compositions of systems thanks to multiple simultaneous roles and the interchangeability of elements. | Collective Behaviours are established when interaction between elements in variable structures is modelled using Meta-Structures. Collective properties are due to coherence between sequences of states adopted by corresponding sequences of various systems, which may exist for any period of time. This Coherence is considered here as being represented by Meta-Structures. |
| Different roles and interchangeability are defined by *fixed* rules, as in equations (9), periodicity or quasi-periodicity, and by using of the same cognitive system | Coherence is due to Meta-Structures |

Table 3: Differences between coherent Multiple Systems/coherent Collective Beings and Collective Behaviours modelled by Meta-Structures

The purpose of the project is to evaluate how values, i.e., meta-elements, assumed to set mesoscopic variables, represent different local stabilities over time whose coherence is described through Meta-Structures and which establish Collective Behaviours. Different local coherent stabilities over time are considered as corresponding to multiple dynamic structures. Such stabilities should be suitably represented by Meta-structural values and their coherence by Meta-Structural properties.

In this generalised view coherent Multiple Systems and coherent Collective Beings are particular cases of Collective Behaviours when considering coherent, *simultaneous* and possibly, at different times, systems, as in Table 2.

Coherence, i.e., Collective properties of coherent Multiple Systems and coherent Collective Beings are considered as arising from continuity and coherence of the multiple dynamic compositions of systems due to the multiple simultaneous roles and interchangeability of elements interacting in different ways as detected by the observer (Minati and Pessa, 2006, pp. 110-143) and due to fixed rules, like equations (9), periodicity and quasi-periodicity of systems, i.e., structures, and role of the *same* cognitive system used by autonomous agents to interact. However, here we consider the coherence of generalised Collective Behaviours as being due to properties of mesoscopic clusters of elements, i.e., Meta-Structures as introduced below. Meta-Structures should represent properties of sequences of states of different individual systems established by the same elements interacting in different ways, i.e., by different structures as sequences of coherent phase transitions. Although in the case of coherent Multiple Systems and coherent Collective Beings the theoretical role of the observer is fundamental for the detection of emergent properties, in the case of Meta-Structures the theoretical role of the observer is fundamental to *invent* a Meta-Structural level of description in order to model the processes of emergence of acquired properties without needing to explicitly specify them, such as variations of ergodicity corresponding to structural changes without specifying the possible new property acquired.

In this case the values of mesoscopic variables consist, for instance, of the *numbers of elements* adopting parameters defining the mesoscopic variable itself, e.g., the *sam*e distance, speed, direction, topological position or altitude of elements over time. Values of mesoscopic variables, parameters defining the mesoscopic variable itself, their relationships and mathematical properties of the sets of values taken over time are assumed as suitable *indices* for *representing* Collective Behaviours just as temperature is a suitable index representing molecular agitation or the *variation of ergodicity* informs the observer of the occurrence of a process of self-organisation or emergence, i.e., a process of re-structuring, even though the observer is unable to detect the corresponding new property being established.

The purpose of the project is to validate the approach in cases where data are easily available and reliable as for computational emergence, e.g., simulated collective behaviours, and then for phenomenological emergence as for the case of industrial Districts.

### 2.1.7 Mesoscopic variables, Meta-elements and Meta-structures

One crucial step of the project relates to the constructivistic (Butts and Brown, 1989) *invention* by the observer of a suitable Mesoscopic level of description. This relates to properties of the cognitive system possessed by the observer, i.e., having specific memory capabilities, image processing, cognitive processing, and input representation capabilities. The cognitive system carries out one or more specific cognitive model(s), i.e., representations using both a level of description given, in short, by the variables, relationships and interactions considered, scalarity, thresholds, time frame, knowledge, space within which possible actions occur and a language representing, explaining and even abstractly, reproducing the phenomenon under consideration. A language of this kind may be related to Gestalt perception and the possibility of describing processes. Bongard proposed a method of creating an adequate language for *visual pattern recognition*, i.e., the use of a language with which the *creation* of an object could be described. His ‘good continuation’ principle – one of the basic principles of Gestalt psychology – assuming that perception of a configuration includes the imaginable process of recreating, or imitating allows the ‘imitation principle’, i.e., imitating the way in which the adopted configuration was created is also an effective way for describing it (Arnheim, 1997; Bongard, 1970). It may help to explain why it is possible to realize that a picture of birds is in reality the picture of a flock: we imagine possible past and future states as being *coherent* with the cognitive models we have in mind, stating the degrees of freedom for the ways in which the configuration may change and keep coherence, i.e., still represent a flock.

Actions and rules effectively used in the world of the observer are used to execute cognitive models. In this view the observer models phenomena by using the possibilities possessed by the observer's cognitive system, the usual or a more appropriate level of description, and a related language describing, explaining, and reproducing the phenomena. In the case where the usage of the usual cognitive model and language is not effective, the observer may constructivistically model the phenomena in different, more adequate ways, i.e., inventing a new model and language. However such invention is performed by using the same cognitive system and by elaborating the models and the knowledge available. This is possible by generalising through a process of abduction, i.e., *invention of hypotheses* introduced by Pierce (Andreewsky and Bourcier, 2000; Peirce, 1998; Von Foerster, 2003). In Section 1 we mentioned the microscopic and macroscopic levels of description used to study Systems and Collective Behaviours. The intermediate, *mesoscopic*, level relates to reduced macroscopic variables without completely ignoring the degrees of freedom at the microscopic level, such as considering in a traffic queue the variable consisting of cars unable to accelerate. In this case the variable relates at the same time to queueing cars not moving, reducing the speed, and having constant speed. The basic idea is to cluster variables in groups having some homogeneity sufficiently significant for modelling the phenomenon considered as in Table 4. This provides some aspects of *continuity* between microscopic and macroscopic properties, and the different language required to model emergent, acquired properties (see Appendix 2).

Although in organised systems it is possible to identify stable subsystems interacting in a suitable structure, Collective Behaviours can be represented as sequences of various coherent systems or coherent states adopted by sequences of systems. Such sequences can be modelled by considering mesoscopic dynamic clusters of suitable mesoscopic variables. Instead of considering structures as in organised systems, mesoscopic regularities of dynamic clusters formally expressed by Meta-Structures are considered.

### 2.1.7.1 Mesoscopic state variables

As a first approach, Mesoscopic state variables represent clusters of agents in Multiple Systems and Collective Beings taking on the *same roles at different times and different roles at the same time*. For instance:

- the same roles at different times: the same values of speed, altitude, direction, distance or topological position define mesoscopic variables at different times;
- different roles at the same time: agents belonging to a mesoscopic variable at a given moment in time, may also belong to another one.

In our generalised approach we no longer consider iteration, replications and the adoption of multiple roles simultaneously or at different times ruled as considered above. As seen below, coherence will be considered as being represented by suitable Meta-Structures. **We consider Mesoscopic state variables relating, for instance, to the *number* of elements $e_k$ having the *same* value** (the observer will consider values as *equal* when *within a range of values*) taken by some specific variables such as the *same* distance from their nearest neighbours, the *same* speed, the *same* direction or the *same* altitude over time, such as:

- $Mx(t_i)$ number of elements having the maximum distance at a given point in time;
- $Mn(t_i)$ number of elements having the minimum distance at a given point in time;
- $N_1(t_i)$ number of elements having the *same* distance from the nearest neighbour at a given point in time;
- $N_2(t_i)$ number of elements having the *same* speed at a given point in time;
- $N_3(t_i)$ number of elements having the *same* direction at a given point in time;
- $N_4(t_i)$ number of elements having the *same* altitude at a given point in time;
- $N_5(t_i)$ number of elements having the *same* topological position at a given point in time

However, $n$-elements constituting a mesoscopic state variable at instant $t_i$ may, in their turn, be clustered into groups having the *same* values as, in the case of distance, $n_1$ are at distance $d_1$, $n_2$ are at distance $d_2$, etc. Thus $n_1 + n_2 + \ldots + n_s$ may be $> n$ (same elements constitute different clusters), $< n$ (not all elements constitute different clusters) or $= n$ (each element belongs to one cluster only).

For this reason sequences of multiple values of a specific mesoscopic state variable over time, should be more properly substituted by sequences of vectors of multiple values. In the same way different sequences of values of different related specific mesoscopic state variables over time consist of sequences of vectors.

For reasons of homogeneity, vectors representing the same Mesoscopic variable are considered as having the same dimensionality assumed to be coincident with the maximum value acquired over the whole period of observation.

For instance, the Mesoscopic vectorial state variable $V_{d,t}$ may be considered as scalar values over time representing the number of elements $e_k$ at the *same* distance $d_q$

$$V_{d,t} = [n_1, n_2, \ldots, n_i] \qquad (10)$$

As stated above, over the whole period of observation all Mesoscopic vectorial state variables related to the *same* parameter, such as distance, are assumed to have the same maximum dimensionality for reasons of homogeneity. Scalar values may be ordered such that $n_i \geq n_{i-1}$.

The simplest case occurs when vectorial dimensions are equal to *1*, i.e., all mesoscopic state variables are defined by a single parametrical value such as one single maximum distance defining the Mesoscopic state variable relating to elements having the same distance at time $t_i$.

It is possible to consider the *dimensions* of Mesoscopic vectorial state variables, *Min* and *Max* of the number of elements $n_i$ for each Mesoscopic vectorial state variable adopted during the whole period of observation as *indices of mesoscopic granularity*.

We underline for future considerations that this level of description is suitable for considering the percentage of elements having the *same* property and the percentage of time spent by elements taking on this property for identifying *mesoscopic ergodicity* at this level of description of the system.

### 2.1.7.2 Meta-elements

Meta-elements are time-ordered sets of values in a discrete temporal representation, *specifying* mesoscopic state variables. While Mesoscopic state variables take as values, for instance, the number(s) of elements having the same property over time, meta-elements are sets of corresponding values taken by that property, such as distance, speed, direction or altitude considered over time.

In the previous Section we defined Mesoscopic vectorial state variable $V_{d,t}$ as given by scalar values over time representing the number of elements $e_k$ which are at the *same* distance $d_q$ with $V_{d,t} = [n_1, n_2, \ldots, n_i]$. The mesoscopic state variable is defined by considering elements sharing the same properties, such as distance, speed, direction or altitude *without specifying values*.

In this Section we correspondingly define Meta-Elements as vectors of values defining the mesoscopic state variables. When considering distance, for example:

$$V_{dis,t} = [d_1, d_2, \ldots, d_i] \qquad (11)$$

Note that $V_{d,t}$ and $V_{dis,t}$ clearly have the same dimensionality *i.*

As stated above, over the whole period of observation we assume that all Mesoscopic vectorial state variables related to the *same* parameter, such as distance, and the vectors of their values, have the same maximum dimensionality for reasons of homogeneity. Scalar values may be ordered such that $d_i \geq d_{i-1}$ or may *correspond* to the order adopted for the related Mesoscopic vectorial state variable $n_i \geq n_{i-1}$. Conversely, the order adopted for the related Mesoscopic vectorial state variable may *correspond* to the order adopted for the related Meta-element $V_{dis,t} = [d_1, d_2, \ldots, d_i]$ such that $d_i \geq d_{i-1}$. Correspondingly, the simplest case occurs when vectorial dimensions are equal to *1*, i.e., all mesoscopic state variables are defined by a single parameter value such that one single maximum distance defines the Mesoscopic state variable relative to elements having same distance at time $t_i$.

### 2.1.7.3 Meta-structures

A Meta-structure is given by the mathematical properties possessed by ordered sets of values adopted over time by Mesoscopic state variables, such as the number of elements having one or more properties and by Meta-elements, such as the values of parameters used to define corresponding Mesoscopic state variables.

Mathematical properties may be statistical, due correlation and auto-correlation, represented by interpolating functions, quasi-periodicity, levels of ergodicity -*meta-structural ergodicity*-, or possible relationships between them in an *N-space*, suitable for modelling a kind of *entropy of correlations*. While structure relates to interacting elements, meta-structures relate to mesoscopic state variables and their parametric Meta-elements. Other examples of Meta-Structures are given by properties of sets of values assumed over time by the mesoscopic *general* vector, see 2.1.8.

In organised systems, Meta-Structures *coincide* with dynamic structural properties.

Meta-elements may be considered as existing in a phase space where each mesoscopic state variable of the system is associated with a coordinate axis. Properties of Collective Behaviour may be represented as the motion of a point along a trajectory within this space. Within this conceptual framework we may apply the usual approaches considered in physics, including the existence of periodic or strange attractors, and fixed points allowing a qualitative analysis.

A tool for research into Meta-Structural mathematical properties is the 'mesoscopic *general* vector introduced' introduced in Section 2.1.8.

Mesoscopic values and Meta-Elements may always exist in Collective Interactions. However, coherent dynamical changes of structures over time are represented by suitable Meta-Structural properties representing the changing of values of mesoscopic variables and their parameters, i.e., Meta-Elements. Sections 2.1.8, 3 and 5 discuss how Meta-Structures may be considered necessary and sufficient conditions for the establishment of Collective Behaviours and the possibility of multiple equivalent or non-equivalent Meta-Structures for modelling Collective Behaviours considered as equivalent when acquiring the same properties, e.g., shape or behaviour.

The approach used is based on considering mesoscopic variables, Meta-elements and Meta-Structures (Minati, 2008a) and is inspired by the concept of *order parameter* introduced in Synergetics when dealing with a mesoscopic level of description (see Appendix 2). In Synergetics the behaviour of a system close to a critical point has been hypothetically considered as being analogous to that of a system undergoing a *phase transition*. This hypothesis, known as the *adiabatic approximation*, implies that the amplitudes of fluctuations in all stable modes can be expressed in terms of the amplitude of fluctuation in only an unstable mode.

Processes of the establishment of coherent Multiple Systems or coherent Collective Beings can be suitably modelled using temporal *successions of degrees of ergodicity* (Minati and Pessa, 2006, pp. 291-319) in order to express the percentage of agents for which there is conceptual *interchangeability* between interacting agents which take on the *same roles at different times and different roles at the same time*, i.e., assuming an ergodic behaviour. This percentage, in turn, could be viewed as a sort of *order parameter*, suitable for describing the degree of progress of the 'phase transition' leading to the formation of the Multiple System or Collective Being itself.

In our approach Collective Behaviours are intended as *coherent sequences* of phase transitions, i.e., states adopted by sequences of different single systems or variable coherent structures (Minati, 2008a). Correspondingly, we conceptually consider *coherent sequences* of possibly simultaneous and different values taken by mesoscopic variables and their properties represented by Meta-Structures. In the case of Multiple Systems and Collective Beings, coherence, between simultaneous or dynamic (different times) systems, is due to multiple roles and an interchangeability of elements ruled by fixed constraints, like equations (9), periodicity and quasi-periodicity of systems, i.e., structures, role of the same cognitive system used by autonomous agents to interact.

Coherence of sequences of multiple instantaneous or individual systems established by the same interchangeable elements interacting in different ways is expected to have a suitable Meta-Structural description once the relative suitable mesoscopic variables have been defined.

### 2.1.8 The mesoscopic *general* vector

For the research outlined above, we propose using a tool to properly find suitable properties. We assume, for instance, that in Simulated Collective Behaviour it is possible to have available at each instant, i.e., computational step, all the required information.

Following the identification of a suitable Mesoscopic Level of Description and during the simulation, at each computational step the researcher has information available about labelled elements belonging to single mesoscopic variables. Suppose Simulated Collective Behaviour occurs in a discretised time of events $t_s$ for *k*-interacting elements. By considering Mesoscopic state variables and Meta-elements taking the simplest case when vectors have dimension *1*, i.e., scalars, we may hypothesize the use of an *m-dimensional* vector $V_t$ representing, for each computational step, how each element satisfies one or more of the *m*-mesoscopic properties used to establish variables:

| | mesoscopic variable$_1$ such as $N_1(t)$ = number of elements with the *same* distance from their nearest neighbour at a given point in time | mesoscopic variable$_2$ such as $N_2(t)$ = number of elements with the *same* speed at a given point in time | mesoscopic variable$_3$ such as $N_3(t)$ = number of elements with the *same* direction at a given point in time | mesoscopic variable$_4$ such as $N_4(t)$ = number of elements with the *same* altitude at a given point in time | mesoscopic variable$_5$ such as $N_5(t_i)$ = number of elements with the *same* topological position at a given point in time | mesoscopic variable$_m$ … |
|---|---|---|---|---|---|---|
| Element $e_1$ | The element takes the value on-off depending on its belonging to $N_1$ | The element takes the value on-off depending on its belonging to $N_2$ | The element takes the value on-off depending on its belonging to $N_3$ | The element takes the value on-off depending on its belonging to $N_4$ | The element takes the value on-off depending on its belonging to $N_5$ | … |
| Element $e_2$ | The element takes the value on-off depending on its belonging to $N_1$ | The element takes the value on-off depending on its belonging to $N_2$ | The element takes the value on-off depending on its belonging to $N_3$ | The element takes the value on-off depending on its belonging to $N_4$ | The element takes the value on-off depending on its belonging to $N_5$ | … |
| ….. | The element takes the value on-off depending on its belonging to $N_1$ | The element takes the value on-off depending on its belonging to $N_2$ | The element takes the value on-off depending on its belonging to $N_3$ | The element takes the value on-off depending on its belonging to $N_4$ | The element takes the value on-off depending on its belonging to $N_5$ | … |
| Element $e_k$ | The element takes the value on-off depending on its belonging to $N_1$ | The element takes the value on-off depending on its belonging to $N_2$ | The element takes the value on-off depending on its belonging to $N_3$ | The element takes the value on-off depending on its belonging to $N_4$ | The element takes the value on-off depending on its belonging to $N_5$ | … |

For instance element-*k* at time *t* may satisfy or not the *m*-mesoscopic property. This is represented by the values *0,1* taken by *k* elements of the vector for each of the *m* mesoscopic variables. For instance:

$$V_{k,t} = [e_{k,1}, e_{k,2}, ..., e_{k,m}] \qquad (12)$$

where:

*k* identifies the element $e_k$;
*t* is the computational step or instant in a discretised time;
*m* identifies one of the *m* mesoscopic properties;
$e_{k,m}$ takes the value = *0* if the element $e_k$ does not have the *m-mesoscopic* property at time *t*;
= *1* if the element $e_k$ does possess the *m-mesoscopic* property at time *t*.

Examples are:

$V_{1,t}$ = [1,1,0,0,1,…,0]
$V_{2,t}$ = [0,1,1,0,0,…,1]
…….
$V_{k,t}$ = [0,0,0,0,0,…,0]

Elements $e_{k,m}$ may only take logical values of on or off, i.e., *0* or *1*. Such values are given on the basis of values defining the mesoscopic level of description such as speed, distance, altitude, topological position or direction.

The same approach works when dealing with Mesoscopic state variables, such as $V_{d,t}$ = $[n_1, n_2, ..., n_i]$ and Meta-elements $V_{dis,t}$ = $[d_1, d_2, ..., d_i]$, as vectors with *same* dimension > 1. We still consider the logic vector $V_{k,t}$ = $[e_{k,1}, e_{k,2}, ..., e_{k,m}]$ by referring to the validity of the mesoscopic property *m* per element *k* over time, *independently* from considering individual parameter values in Meta-elements, i.e., $e_{k,m}$ is considered *on* (equal to 1) if *any* of the $n_i$ constituting the Mesoscopic state variable $V_{d,t}$ is different from zero.

By considering *sequences* over time of vectors $V_t$ it is possible to detect, for instance:

*Continuity*

1) How many and which elements have the same or different or no mesoscopic properties over time;
2) How many and which elements have more than one mesoscopic property over time;
3) Multiple repetitiveness, coherence and local, partial quasi-periodicity;
4) Topological distance between elements having a mesoscopic property.

The *set* of all vectors $V_{t:1,s}$ allows one to detect, for instance:

*Diffusivity*

1) Number of computational steps between states on-off per element with regard to all mesoscopic properties;
2) Number of computational steps occurring before *all elements* have taken at least once the state *on* (indicated as *general meso-state on*), repetitiveness;
3) How many times the *general meso-state on* occurs, i.e., how many times it takes the state *on*;
4) Properties of the sets of numbers of steps: statistical, periodic, quasi-periodic, etc.

Properties of *Continuity* and *Diffusivity* at the Mesoscopic Level of Description are introduced as possible *necessary conditions* for the establishment of Collective Behaviours by collectively interacting elements (see Section 5, point b).

### 2.1.9 The project

The purpose of the project is to find Meta-Structures, i.e., relationships and properties of time-ordered sets of values adopted over time by mesoscopic state variables and parametric Meta-structures, hypothesized *corresponding* and *representing* Collective Behaviours when eventually established. Stability and coherence of variable structures, i.e., acquisition of Collective Behaviour, are assumed as being suitably represented by Meta-Structural properties as outlined in 2.1.7. The prospective advantage on sub-symbolic approaches is the conceptual possibility to *induce* or *prescribe* the adoption of Meta-Structures in Collective Interactions and to consider Meta-Structure as representing specific kinds of Collective Behaviours.

The first step of the *Meta-Structures project* relates to the possibility of detecting suitable Meta-Structures in Simulated Collective Behaviours following the identification by the observer of one or more suitable Mesoscopic Levels of Description. Simulation must make collective interacting agents establish a Collective Behaviour detectable by the observer, i.e., acquisition of emergent properties due, in this case, to computational emergence. The main purpose of the project is to use Meta-Structures to model *general processes of emergence* (Minati, 2008a; 2009a). As already specified the approach is *compatible* with the heterogeneous hypothesis when collectively interacting elements, individually labelled, behave within mesoscopic degrees of freedom, following meta-structural constraints as defined by the Meta-Structures introduced above. In Simulated Collective Behaviours, the behaviour of single elements are computed through (possibly *evolutionary*) rules of interaction considering the general configuration, random changes, variations of values taken by some variables such as speed, direction and *visibility* for each elements. For possible topological rather than metrical models see, for instance, (Ballerini *et al.*, 2008). Simulated Collective Behaviours used in this research are *general*, in the sense that we do not claim to simulate any *specific* natural Collective Behaviour. We use, in a general way, the concept of a flock of boids, flock being intended as an emergent entity acquiring computational emergent properties and the boids collectively interacting as general elements. It is possible to consider, for instance, prototype-models to simulate a Collective Behaviour, randomly perturbed, for studying Meta-Structures in the computational emergence of Cellular Automata of Class II (which evolves to simple periodic or quasi-periodic patterns) or Class III (chaotic aperiodic patterns) in the Wolfram-Langton classification (Langton, 1990).

## 3. RESEARCH ISSUES AND APPROACHES

The purpose of the project is to establish a suitable theoretical framework for a possible General Theory of Emergence (Minati, 2008a).

The first line of research relates to the possibility of *demonstrating*:

1) That in any Simulated Collective Behaviour a Meta-Structure *always* exists. A subsequent problem is a *generalisation* to any kind of Collective Behaviour.
2) If and when a Simulated Collective Behaviour has different, *non-equivalent*, i.e., based, for instance, upon different Mesoscopic Levels of Description, Meta-Structure representations.
3) If and when a specific Meta-Structure may represent more, *non-equivalent* Simulated Collective Behaviours, i.e., consisting, for instance, of a different number of elements, rules of interaction and scalarity.

A subsequent step relates to the possibility of detecting *similar* meta-structures in different Simulated Collective Behaviours having, for instance, different rules of interaction between elements, timeframes, scalarities, or numbers of elements. Two Meta-Structures may be considered *similar* when, for instance:

a) They use the same Mesoscopic Level of Description with *m*-variables or Mesoscopic Levels of Description differing by $\delta < m$ mesoscopic variables considered. The number $\delta$ of different variables considered specifies the *mesoscopic granularity* assumed.
b) Meta-structural properties representing two different Simulated Collective Behaviours at the same Mesoscopic Level of Description may differ by a number $\eta$ of properties. The number $\eta$ of different properties considered specifies the *metastructural granularity*.

Another subsequent area of research relates to possible *correspondences* between properties of Simulated Collective Behaviours and the corresponding Meta-Structures. The system of correspondences could be taken to represent the phenomenon of Collective Behaviour in general and be studied *per se*, in an even more abstract, i.e., trans-disciplinary way.

The second line of research regards the existence of a Simulated Collective Behaviour or Collective Behaviour for *any possible* Meta-Structure. Within this conceptual framework further additional constraints, if any, must be studied to be added to a *generic* Meta-Structure in order to *actually* represent a Simulated Collective Behaviour or Collective Behaviour.

This research will also allow the possibility to prescribe a Meta-Structure to:

1) Simulated Collective Behaviours in order to *modify* the Collective Behaviour currently adopted with its related acquired properties;
2) Elements collectively interacting without establishing Collective Behaviour, i.e., the observer does not detect modelled processes of acquisition of properties.

## 4. *DYNAMIC* MESOSCOPIC LEVELS OF DESCRIPTION

The Mesoscopic Level of Description should also be assumed to be dynamic, when the observer dynamically *focuses* upon the phenomenon. Meta-Structural modelling is based on a specific

Mesoscopic Level of Description decided upon by the observer. However, this modelling may be dynamic in different ways. It is possible, for instance, to:

1) Identify *domains* of validity for different Meta-Structural properties over time to better model the *changing* of acquired emergent properties or the occurrence of new situations such as variations in environmental conditions. Different Meta-Structural properties may be *valid* at different times. Processes of change between domains, i.e., between related Meta-Structures, should be, in their turn, also modelled.
2) Change the Mesoscopic Level of Description used and have not only new Meta-Structural properties at the *same* Mesoscopic Level of Description as in the previous case, but also have *new* Mesoscopic Levels of Description. They may be simultaneous different Mesoscopic Levels of Description of the same phenomenon, since they represent dynamically the adoption of the *same* emergent property as detected by the observer. However, is it possible to *formally* represent any correspondence between such different levels of description?

Is it possible to consider a dynamic usage of Meta-Structures and Mesoscopic Levels of Description as introduced in Section 6.2?

## 5. A CONCEPTUAL FRAMEWORK FOR ESTABLISHING NECESSARY AND SUFFICIENT META-STRUCTURAL CONDITIONS

A Meta-Structure at a specific Mesoscopic Level of Description establishes the mesoscopic degrees of freedom, i.e., the degrees of freedom for mesoscopic variables.

Can different Simulated Collective Behaviours correspond to the *same* Meta-Structure at the *same* Mesoscopic Level of Description?

This project aims to show that respecting a suitable Meta-Structure at a specific Mesoscopic Level of Description ensures *coherence*, i.e., the acquisition of the emergent property of coherence.

However, can a specific Simulated Collective Behaviour be described by different Multiple Meta-Structures each corresponding to different Mesoscopic Levels of Description? Can two Simulated Collective Behaviours be considered *different* when corresponding to different Multiple Meta-Structures at the same Mesoscopic Level of Description? At this point two different conceptual strategies seem possible:

a) Demonstrating two fundamental theorems about Simulated Collective Behaviours, such as:

Theorem 1 (*Necessary and Sufficient conditions*):

Sufficient condition: The adoption of a Meta-Structure is a sufficient condition for the establishment of Simulated Collective Behaviour by collectively interacting elements.

Necessary condition: Each Simulated Collective Behaviour possesses at least a Meta-Structure at a suitable Mesoscopic Level of Description.

Theorem 2 (*Existence and Uniqueness*):

Existence: At a specific Mesoscopic Level of Description, a specific Simulated Collective Behaviour always has a corresponding Meta-Structure.

Uniqueness: At a specific Mesoscopic Level of Description and by assuming suitable boundary conditions, only one Meta-Structure exists for a specific Simulated Collective Behaviour.

b) *Assume* the existence of Meta-Structural properties as a *definition* of Simulated Collective Behaviour.

General Definition: A collective entity established by *N* elements *collectively interacting* is said to establish, at a specific level of description, a Collective Behaviour when elements collectively interact by respecting one or various Meta-Structures at one specific or at different Mesoscopic Levels of Description. This conceptually *corresponds* to the fact that in infinite state quantum systems studied by Quantum Field Theory different and not unitarily equivalent representations of the same system are possible.

The alternative strategy to that in point a) is to show the existence of Simulated Collective Behaviours for which it is possible to demonstrate the *non-existence* or *contradictoriness* of corresponding Meta-Structures at *any* Mesoscopic Level of Description.

The definition introduced can be considered to be based upon the *absurdity* of negating:

*Necessary conditions* (outlined in Section 2.1.8) including:

1) Properties of *continuity* when elements have more than one mesoscopic property over time;
2) Properties of *diffusivity* when mesoscopic properties are adopted with dynamic and coherent homogeneity.

*Sufficient conditions*, such as:

1) The existence of Meta-Structures, i.e., properties of sets of values of meta-elements;
2) *Analyticity* of Meta-Structures. A phase of a physical system is defined in physics as a region in the parameter space of the system's thermodynamic variables where the *free energy* is *analytic*. This means that in a region within the parameter space of the system's thermodynamic variables the *free energy* can be transformed in an *analytic* way, i.e., the transforming function is infinitely differentiable and can be described by a Taylor series. In the same we may require values of Meta-Structures be *analytic*. Is it also possible to consider Meta-Structures as representing phases of matter or phase transitions?

## 6. META-STRUCTURES, BOUNDARY CONDITIONS AND META-STRUCTURAL DYSAM

### 6.1 Boundary Conditions and Meta-Structures: the case of Architecture structuring space

We use the term Boundary Conditions, from the original meaning acquired when considering infinite solutions of differential equations, to refer to *partial microscopic* or *macroscopic degrees* of freedom for interacting elements. Boundary Conditions lead to the acquisition of macroscopic properties such as the dimensions and shapes of pipes which *induce* the emergence of whirlpools from the behaviour of interacting molecules in fluidodynamics. Similarly width, shape, visibility, ascents and descents allow the emergence of properties for traffic and crowds.

From this viewpoint, architectural structures within a given space, as for houses, hospitals and schools, may also be considered as Boundary Conditions representing partial behavioural constraints for inhabitant agents. For references to the concept of the self-architecture of social systems, see Minati (2009b) and Minati and Collen (2009). This generalised concept of Boundary Conditions may be also used as a *methodology*, for instance through simulation, to anticipate, prescribe and change behaviours acquired by elements collectively interacting by respecting such specific Boundary Conditions. In architecture the topic relates to *Pre-Occupancy Evaluation* in alternative to the usual *Post-Occupancy Evaluation* (POE) as in Oseland (2007).

Is it possible to *transform*, even partially, Meta-Structures into suitable, even dynamic, Boundary Conditions? This will eventually make it possible to indirectly *prescribe* Meta-Structures to Collective Interacting elements in a *non-invasive* way, i.e., acting upon environmental properties without acting upon elements or interactions. Suitable theorems of equivalence should be demonstrated.

Boundary Conditions and Meta-Structures may then also be considered as constraints for modelling and influencing multiple interactions in Multiple Systems and Collective Beings.

### 6.2 Meta-Structural DYSAM

The Dynamic Usage of Models (DYSAM) approach (Minati and Pessa, 2006, pp. 64-88; Minati and Brahms, 2002) was introduced to deal with subsequent processes of the acquisition of emergent properties in complex systems, coherent Multiple Systems and coherent Collective Beings.

What is DYSAM? In order to model a Multiple System we need different representations related to each component system over time. Multiple different levels of description and models are resources for multiple representing and modelling Multiple Systems and Collective Beings.

DYSAM is a meta-model, i.e., a model of models having in its turn dynamic systemic properties. In short, the main components of DYSAM are:

- *a repertoire of different possible models* of the same system;
- *a strategy* for selecting and deciding, on the basis of general and momentary goals, the available knowledge and the context, the models to be used (and eventually integrated) to model the system considered from simultaneous different approaches;
- this strategy is not only variable, being based, for instance, upon learning (and not*only optimisation*), but on redefining and modelling interactions between the adopted models. Moreover, not only is the strategy variable, but evolutionary as it varies with the evolution of the interactions between the observer and the system.

An implementation of DYSAM based on Neural Networks was introduced by Minati and Pessa (2006, pp. 75-85).

DYSAM deals with the *usage* of multiple models and their relationships of corresponding sequences of multiple fixed structures, i.e., Multiple Systems and Collective Beings, relating to the process of acquiring emergent properties in a complex system. This usage requires models for the processes of emergence, naturally available, still not explicit, in sophisticated cognitive processes performed, for instance, by human beings.

The purpose of DYSAM is to use in the most effective way several possible levels of description and models to deal with a complex phenomenon, where usage of a single model is in principle ineffective due to processes of emergence of new acquired properties. Some examples are different levels of description and models used in disciplines, like quantum or non-quantum physics, sociological or economical in social systems, biological or psychological in human care, deciding between possible corporate strategies, usage of the system of remaining resources in a damaged system, e.g. acquired disabilities, and learning the usage of the five sensory modalities in the evolutionary age where the purpose is not to select the best one but to use all of them together in a coherent way making behaviour possible. DYSAM is based on putting ‘and’ between available resource and using them in coherent way rather than ‘or’ like in non systemic approaches.

However, DYSAM may be assumed to deal not only with models using microscopic and macroscopic levels of description, but also with models using the mesoscopic approach. We may consider *transitions between Meta-Structures* corresponding to processes of transition in Collective Behaviours from one acquired property to another one non-suitably modelled by the previous Meta-Structure. In this case DYSAM deals with different Meta-Structures over time to cope with dynamic *domains* of validity as introduced in Section 4. With reference to the concepts of Quantum Field Theory this corresponds to processes in which phase transitions structurally modify the system through *Spontaneous Symmetry Breaking* leading to not all the states being compatible with a given energy value invariant.

## 7. FUTURE LINES OF RESEARCH AND POSSIBLE APPLICATIONS.

### 7.1 Future lines of research

Future lines of research relate to both experimental and theoretical issues.

a) Experimental issues

Experimental issues relate to the possibility of handling non-simulated phenomena of Collective Behaviours where it is possible to easily have a suitable measure of values with which to search for Meta-Structures and run suitable models, as in socio-economic systems, e.g., Industrial Districts. Other experimental issues, mentioned above, relate to the possibility of inducing Collective Interacting elements to behave according to a Meta-Structure through the usage of suitable, corresponding, Boundary Conditions or by suitable perturbations such as interaction with micro-Collective Behaviours respecting the Meta-Structure we want to induce.

b) Theoretical issues

Theoretical issues relate to the possibility of:

- considering Meta-Structural research not only in Collective Behaviours established by interacting elements at a given time $t_i$ as for flocks, but also in *sequences* of configurations of correlated even non-interacting elements such as sequences of points marking a face over time, e.g., allowing face recognition, or notes in music or words in a written text. In this case Meta-Structural research does not have the purpose of modelling the emergence of Collective Behaviour from collectively interacting elements, but rather to model the emergence of collective properties from sequences of sets of configurations of component and even non-interacting elements. In the first case, the emergent property acquired by the collectively behaving elements is conserved over time by coherent collective behaviour adopted by the elements, e.g., a flock of boids, and Meta-Structures have the purpose of modelling this process. In the second case, we consider *sequences* of configurations of values taken by variables over time. *The focus is not placed upon relations, nor upon interactions between variables, but rather upon the properties of sequences*. This the difference between a flock and a smiling face over time;
- *analytically* identifying, using formal models and computational processes, such as adiabatic reductions, suitable Mesoscopic Levels of Description and correspondences between Collective Behaviours and Meta-Structures. In this way mesoscopic variables may be no longer considered as being carried out only by the observer;
- considering Meta-Structures not only between *elements* as particles, but, for instance, between quasi-particles and processes. "… what level of fluctuations can we tolerate in order to consider a particle as an almost invariant entity, endowed with a specific identity? Could we identify, for instance, a particle with some kind of statistical construct related to empirical data? Is the concept of particle a fuzzy concept? Could we, up to a certain degree, deal with particles in the same way as zoologists deal with animal species?" (Pessa, 2009). This relates to considering particles as *probabilities* and considering Meta-Structures between *probabilities* allowing further generalisations including models for quantistic entanglement;
- The *Renormalization Group* studied in physics (Gallavotti, 1985; Shirkov and Kovalev, 2001) has been very helpful in giving a solid theoretical basis to processes of phase transition phenomena and transforms every phase transition problem into a field-theoretical problem and vice versa. It is a powerful mathematical mechanism to identify the ‘correct’ and physically ‘stable’ phenomenological scales. Is a similar approach possible for Meta-Structures to identify the proper scale where values of parameters such as speeds, distances, altitudes and granularities are considered *equal* when they lie within a range of values and identify the most effective mesoscopic level of description?

- exploring possible theoretical relationships between so-called *Fisher information* (Frieden, 2004) and Meta-Structures;
- exploring possible theoretical relationships between *quantum emergence* (see, for instance, Licata and Sakaji) and Meta-Structures.

### 7.2 Possible applications

a) Image recognition and understanding

Possible areas of application relate to situations where macroscopic properties cannot be suitably reduced to relationships between elements. Typical fields of interest are image understanding and face recognition when considered as properties acquired by Collective Behaviours of stimuli acquired over time. Meta-Structures may be studied as representing the meaning of Collective Behaviours of stimuli over time intended as elements, considered as images and allowing recognition. Due to the nature of Mesoscopic Levels of Description and Meta-Structures this possible approach seems more suitable when phenomena are dynamic rather than static, e.g., video sequences.

We may hypothesize that as we are able to detect coherence in collective phenomena, we may use the same detection system to detect coherence amongst various properties only superficially assumed as being non-collective such as image understanding. Actually, they are detected through binding processes, which are processes of coherence and properties of Collective Behaviours established by collective populations of stimuli and neurons. It is possible to apply Meta-Structures and we may hypothesize the availability of devices able to detect Meta-Structures associated with meanings. This may be due a) to learning and b) to some assumed embedded parameter values as for the famous *Grounding Problem*, consisting of the fact that a neural network is unable to connect higher-level symbolic representations with lower-level space-time distributions of physical signals coming from the environment.

b) Development in socio-economic systems

It is possible to consider the emergence of processes of development from collectively interacting processes of growth in socio-economic systems (Minati and Pessa, 2006) as modelled by suitable Meta-Structures. Coherence between processes of growing is not reduced to harmonic relations, but considered as being due to Meta-Structures.

c) Transforming the needs of biological living matter into behaviour

Suitable Meta-Structures *detected* in systems of inputs such as visual, acoustic, olfactory, or tactile signals, may *model*, if not *explain*, the processes of transforming the basic needs of biological living matter, such as sexual activities, the search for food, or the protection of the new-born, into cognitive properties, see the concept of BioCognitiveConverter (BIOCC) in Minati (2008a; 2009a).

## CONCLUSIONS

We have presented the current state of the art of the Meta-Structures project, presenting critical research issues, open issues and a formal tool, (the mesoscopic *general* vector), for representing and modelling Meta-structural properties of a Collective Behaviour. We presented general principles at the basis of the approach including the use of variable structures, the previously introduced concepts of Multiple Systems, order parameters in Synergetics and considering Collective

Behaviours as *coherent sequences* of phase transitions, i.e., variable coherent structures. We have conceptually considered *coherent sequences* of states acquired by various single systems established by the same elements interacting in different ways, i.e., different structures, as related to the values taken by Mesoscopic state variables and their properties represented by Meta-Structures. We have also introduced possible relationships between Meta-Structures and Boundary Conditions and some possible approaches towoards the use of dynamic Boundary Conditions as transforms of Meta-Structures to influence the emergence of Collective Behaviours. We have mentioned the possibility of introducing a Meta-Structural version of DYSAM and to use Meta-Structural changes to model the processes of the acquisition of new properties in Collective Behaviours.

We have also presented future lines for theoretical research and the possibility of applying Meta-Structures to collective populations of different natures, such as neurons and stimuli, to model processes of image understanding and the binding problem. This may also be a suitable approach to the BIOCC.

## APPENDIX 1: EXAMPLES OF RESEARCH FOR FINDING META-STRUCTURES IN SIMULATED COLLECTIVE BEHAVIOURS

a) Properties of Mesoscopic variables

Consider a Simulated Collective Behaviour with a fixed number of boids over the simulation time $t_{:1\text{-}s}$ where $s$ is a finite and limited number of computational steps. The mesoscopic *general* vector $V_{k,t}$ = [$e_{k,1}$ , $e_{k,2}$ , …, $e_{k,m}$] represents at each instant which elements have one or more of the *m-mesoscopic* properties considered at the Mesoscopic Level of Description adopted.
This allows one to compute at each instant, for instance:

1) $y_1$ - number of elements having a specific mesoscopic property;
2) $y_2$ - number of elements having more than one mesoscopic property;
3) $y_3$ - number of elements having maximum distance, speed or altitude;
4) $y_4$ - number of elements having the minimum distance, speed or altitude;
5) $y_5$ - number of computational steps between on-off states per element and regarding all mesoscopic properties;
6) $y_6$ - number of computational steps occurring before *all elements* have acquired at least once the *on* state (indicated as a *general meso-state on*);
7) $y_7$ - number of times the general meso-state is *on*;
8) positional properties of elements allow one to model the *spatial dispersion* of a mesoscopic property;
9) positional properties of elements also allow one to model eventual topological aspects related to elements having a mesoscopic property;

It is also possible to consider the percentage of elements having the *same* property and the percentage of time spent by elements actually possessing this property to identify eventual *mesoscopic ergodicity* at this level of description of the system.

b) Meta-Structural properties

We recall that meta-elements are time-ordered sets of values in a discrete temporal representation, *specifying* mesoscopic variables.
While mesoscopic state variables take as their values, for instance, the number of elements having the same property over time, meta-elements are sets of corresponding values taken by a property, such as distance, speed, altitude or direction, considered over time. Examples of meta-elements at each instant $t$ are given by considering over time the sets of values specifying mesoscopic state variables, as for the mesoscopic general vector introduced in Section 2.1.7:

10) $y_8$ - value of the maximum distance, speed or altitude at time $t$;
11) $y_9$ - value of the minimum distance, speed or altitude at time $t$;
12) $y_{10}$ - value of the distance, considered suitable for specifying a mesoscopic variable at time $t$, e.g., max or min or weighted average among those corresponding to different simultaneous clusters considered in the vectorial description;
13) $y_{11}$ - value of the speed considered suitable for specifying the mesoscopic variable at time $t$, e.g., max or min or weighted average among those corresponding to different simultaneous clusters considered in the vectorial description;
14) $y_{12}$ - value of the direction considered suitable specifying the mesoscopic variable at time $t$, e.g., max or min or weighted average among those corresponding to different simultaneous clusters considered in the vectorial description;
15) $y_{13}$ - value of the altitude considered suitable for specifying the mesoscopic variable at time $t$, e.g., max or min or weighted average among those corresponding to different simultaneous clusters considered in the vectorial description.

Previous values may be considered over the total observational time, i.e., for the *set* of all vectors $V_{t:1,s}$ , allowing the plotting over time of all the values listed above and identifying eventual regularities and cross-correlations. Meta-structures also consist of other possible properties as mentioned in Section 2.1.6.3.

## APPENDIX 2: MICROSCOPIC, MACROSCOPIC AND MESOSCOPIC APPROACHES

We prefer not to refer to a *spatial dimensional scale* of considered elements, but rather to representations in terms of properties acquired. The term *microscopic* relates to *single* elements, such as the position or speed of particles, cars or planets; *macroscopic* relates to properties not reducible to those of microscopic variables, such as pressure, temperature or financial indices; and the intermediate level, known as *mesoscopic*, relates to reduced macroscopic variables without completely ignoring the degrees of freedom of the microscopic level, as one considers the variable consisting of cars unable to accelerate in a traffic queue. In this case the variable relates at the same time to cars in a queue not moving, reducing their speed, or having a constant speed.

| | |
|---|---|
| **Microscopic level** | Structures *completely* specify, with reference to some considered microscopic state variables, the way in which elements interact. Other variables are considered insignificant for the description of system behaviour. Examples are given by equations of mechanics and gravitation specifying the movement of bodies in physics, circuit board in electronics specifying interaction among components when supplied with power and Agent-Based Models. |
| **Macroscopic level** | Structures *completely* specify how some considered macroscopic state variables interact. Microscopic variables have the freedom to behave in *any* way, but respecting macroscopic constraints as defined for macroscopic state variables. Other variables are considered insignificant for the description of system behaviour. Example: deterministic equations of physics, as in thermodynamics or fluidodynamics, model the behaviour of macroscopic state variables such as pressure or temperature; Lotka-Volterra equations do not consider the colour or the sex of elements |
| **Mesoscopic level** | The approach used here is inspired by one of the many important new ideas introduced by Synergetics, i.e., the concept of *order parameter*. When complex systems undergo phase transitions, a special type of ordering occurs at the microscopic level. Instead of addressing each of a very large number of atoms of a complex system, Haken in1988 showed, mathematically, that it is possible to address their fundamental *modes* by means of *order parameters*. The very important mathematical result obtained using this approach consists in drastically lowering the number of degrees of freedom to only a few parameters. Complex systems organize and generate themselves at far-from-equilibrium conditions: *"In general just a few collective modes become unstable and serve as 'order parameters' which describe the macroscopic pattern. At the same time the macroscopic variables, i.e., the order parameters, govern the behaviour of the microscopic parts by the 'slaving principle'. In this way, the occurrence of order parameters and their ability to enslave allows the system to find its own structure".* (Graham and Haken, 1969, p. 13). "*In general, the behaviour of the total system is governed by only a few order parameters that prescribe the newly evolving order of the system*" (Haken, 1987), p. 425. In our approach Collective Behaviours are intended as *coherent sequences* of phase transitions, i.e., variable coherent structures (Minati, 2008a). Correspondingly, we conceptually consider in our approach *coherent sequences* of possibly simultaneous states of various individual systems established by the same elements interacting in different ways as modelled by mesoscopic variables and their properties represented by Meta-Structures. Meta-Structures relate to relationships between mesoscopic variables and properties of the sets of their parameters over time. Mesoscopic state variables are suitable constructivist creations of the observer. Microscopic and macroscopic variables have the freedom to behave in *any* way, but always respecting meta-structural constraints as defined for the mesoscopic state variables. |

## APPENDIX 3: FROM COLLECTIVE INTERACTION TO COLLECTIVE BEHAVIOURS AS DYNAMIC STRUCTURES

| | |
|---|---|
| **Collective Interaction** | Elements are considered as particles and modelled as *all* provided with the *same* behavioural rules, fixed over time. The *high-number* of cases due to 1) elements interacting after starting from different initial conditions, 2) frequencies, combinations, intensity, angularities of random sequences or eventually simultaneous, i.e., parallel, input, 3) different reacting rules, such as gravitational or electrostatic, 4) and different environmental situations, often simulated by random parameters, do not allow classic analytic representations, see Section 1.2. |
| **Collective Behaviour** | If Collective Interaction establishes a collective entity detected by the observer, at a suitable level of description, having properties which elements do not possess, we have processes of emergence of Collective Behaviour. The process is classically intended as being given by the rules of interaction between elements. Equations modelling the evolution of Dynamical Systems, e.g., the Brusselator, Lorenz or Lotka-Volterra equations, may be considered as *suitable structures setting relationships upon macroscopic state variables.* Rules used for evolving, rule-based systems organised in this way, e.g., Cellular Automata or Agent-Based Models, may be considered as specific structures when using specific rules, setting constraints upon elementary interacting elements, i.e., cells and agents. In these cases structures between microscopic or macroscopic variables are assumed as *fixed*, i.e., the analytic representations or rules do not change, see Section 2.1.1. |
| **Multiple Systems and Collective Beings** | A Multiple System is a set of simultaneous or successive systems modelled by the observer and established by the same elements interacting in different ways, i.e., having multiple roles simultaneously or at different times. Successive systems consisting of the same elements interacting in different ways over time and establishing a Multiple System may exist for any period of time. Coherent Multiple Systems are established, for instance, by *interchangeability* between interacting agents to model emergent behaviour as ergodic, and when suitable constraints govern a simultaneous validity of various interactions such as those expressed by equations (9). In the case of self-organisation the coherence of sequences of Multiple Systems is due, for instance, to their periodicity or quasi-periodicity of systems, i.e., structures, such as for Bènard rollers, Belousov-Zhabotinsky reactions, the laser, repetitive collective behaviours, e.g., swarms, and dissipative structures such as whirlpools, when non-perturbed. The persistence of coherence corresponds to a stability of the acquired property. Collective Beings are particular cases of Multiple Systems when elements are provided with same cognitive system, see Section 2.1.5. In the case of Collective Beings coherence between multiple systems is due to the *same cognitive* system and same physical characteristics used by autonomous agents to interact, as when belonging to the *same biological species*. This is the case for flocks, fish schools and swarms. Coherent Multiple Systems and coherent Collective Beings are particular cases of Collective Behaviours. Processes of the establishment of coherent Multiple Systems or coherent Collective Beings have been proposed to be suitably modelled by using the *degrees of ergodicity* to express the percentage of agents for which there is conceptual interchangeability between interacting agents which take on the *same roles at different times and different roles at the same time*, i.e., ergodic behaviour. This percentage, in turn, could be viewed as a sort of *order parameter*, suitable for describing the degree of progress of the 'phase transition' leading to the formation of the coherent Multiple System or coherent Collective Being itself. Coherence of sequences of multiple instantaneous or single systems established by the same interchangeable elements interacting in different ways is expected to have a suitable Meta-Structural description once related suitable mesoscopic variables have been defined. |
| **Collective Behaviours as Dynamic Structures** | In Collective Interaction processes of the emergence of Collective Behaviour are considered as corresponding to continuous, irregular but *coherent*, i.e., as detected by the observer, variability in the acquisition of new structures, as for swarms, flocks, traffic and Industrial Districts adopting variable behaviours when in the presence of suitable environmental conditions. Subsequent systems are not hierarchical, but sequential and coherent over time, i.e., they display to the observer the *same* emergent property. This is the case for the emergence of Collective Behaviours. In this case emergent macroscopic acquired properties are modelled as generated by corresponding systems of collective microscopic interactions. The theoretical role of the observer is to adopt a level of description suitable for detecting emergent properties, see Section 1.2. Values of mesoscopic variables, parameter meta-element values defining the mesoscopic variables themselves, their relationships and mathematical properties of the sets of values adopted over time are taken as suitable *indices representing* Collective Behaviours, as temperature is a suitable index representing molecular agitation and *variation of ergodicity* informs the observer that a process of self-organisation or emergence, i.e., a process of re-structuring, is occurring even without detecting the corresponding new property being established, see Section 2.1.6. Collective Behaviours are intended as *coherent sequences* of phase transitions, i.e., states adopted by sequences of systems as variable coherent structures. In the case of Multiple Systems and Collective Beings, coherence between simultaneous or dynamic (different times) systems, is due to multiple dynamic compositions of systems due to the multiple simultaneous roles and the interchangeability of elements. Correspondingly, here, we conceptually consider *coherent sequences* of possibly simultaneous states adopted by various individual systems established by the same elements interacting in different ways. This can be suitably modelled using the different values taken by mesoscopic variables and their properties represented by Meta-Structures, see Section 2.1.7.3.<br>1. *Indices of mesoscopic granularity*: it is possible to consider the *dimensions* of Mesoscopic vectorial state variables, *Min* and *Max* of a number of elements $n_i$ for each Mesoscopic vectorial state variable taken over the whole period of observation as *indices of mesoscopic granularity*.<br>2. *Mesoscopic granularity*: when using the same Mesoscopic Level of Description or Mesoscopic Levels of Description differing by $\delta$ mesoscopic variables considered, the number $\delta$ specifies the *mesoscopic granularity* assumed.<br>3. *Metastructural granularity*: Meta-structural properties representing two different Simulated Collective Behaviours at the same Mesoscopic Level of Description may differ by a number $\eta$ of properties. The number $\eta$ of different properties considered specifies the *metastructural granularity* assumed. |